# MENSA$_{DB}$: A Thorough Structural Analysis of Membrane Protein Dimers

Pedro Matos-Filipe[1#], António J. Preto[1#], Panagiotis I. Koukos[2], Joana Mourão[1,3], Alexandre M.J.J. Bonvin[2], Irina S. Moreira[1,3*]

[1]*Centro de Neurociências e Biologia Celular, UC-Biotech Parque Tecnológico de Cantanhede, Núcleo 04, Lote B, 3060-197 Cantanhede, Portugal.*

[2]*Bijvoet Center for Biomolecular Research, Faculty of Science – Chemistry, Utrecht University, Utrecht, 3584CH, The Netherlands.*

[3]*Institute for Interdisciplinary Research, University of Coimbra.*

[#]*co-first authors*

***\* Correspondence:***
*Corresponding Author*
*irina.moreira@cnc.uc.pt*

## INTRODUCTION

Membrane Proteins (MPs) account for around 15-39% of the human proteome (Almén et al., 2009; Michael Gromiha and Ou, 2014). They assume a critical role in a vast set of cellular and physiological mechanisms, including molecular transport, nutrient uptake, toxin and waste product clearance, respiration, and signaling (Perez-Aguilar and Saven, 2012). While roughly 60% of all FDA-approved drugs target MPs, there is a shortage of structural and biochemical data on them mainly hindered by their localization in the lipid bilayer (Overington et al., 2006; Yıldırım et al., 2007). In the last few years, a primary goal of drug discovery has been the development of small compounds that can target specifically membrane Protein-Protein Interactions (PPIs) (Feng et al., 2017; Yin and Flynn, 2016). In this respect, being able to characterize structural and physico-chemical properties of MPs and their interactions is essential to develop improved and more targeted therapies, as well as to discover new drug targets.

Some studies working with particular features of proteins, such as electrostatic forces (Zhang et al., 2011), hydrophobic effects (Chanphai et al., 2015) or "hot-spot" residues (Darnell et al., 2008; Moreira et al., 2007, 2017; Rosell and Fernández-Recio, 2018) showed that they could have a contribution in the improvement of the affinity and specificity of PPIs. Another well-characterized property of proteins is the evolutionary conservation and distribution of their amino-acids, which make the most substantial contribution to predicting functionally essential residues as highlighted by several publications (Caffrey, 2004; Capra and Singh, 2007; Ulmschneider and Sansom, 2001; Zhang et al., 2010). Unfortunately, a significant bottleneck is the lack of in-depth analysis of membrane protein complexes and their interactions.

We present here MEmbrane protein dimer Novel Structure Analyser database (MENSAdb), a real time web-application exposing a broad array of fundamental features about MPs surface and their interfacial regions. Users can easily access a thorough, systematic analysis of sequence-structure relationships (**Figure 1**) based on a curated database of 201 protein dimers obtained from the Membrane Proteins of Known 3D structure (MPSTRUCT) (White, 2009).

MENSAdb delivers tabular and graphical data formats that can be visually explored for conservation, four distinctive Accessible Solvent Area (ASA) descriptors, average and environment-specific *B*-factors, intermolecular contacts at 2.5 Å and 4.0 Å distance cutoffs, salt-bridges, hydrogen-bonds, hydrophobic, π-π interactions, t-stacking and cation-π interactions. Additionally, users can closely inspect differences in values between three distinctive residues classes: i) non-surface, ii) surface and non-interfacial and iii) interfacial. It relies on a custom frontend application that provides the results to the user. The resulting knowledge and full datasets can be easily assessed and downloaded. The database is freely available at www.moreiralab.com/resources/mensadb.

## EXPERIMENTAL DESIGN, MATERIALS AND METHODS

### Data collection and pre-processing

Experimental structures of 167 unique transmembrane proteins, that included monotropic MPs, β-barrel TMs and α-helix TMs, were obtained from MPSTRUC – Membrane Proteins of Known 3D Structure (acquisition in September 2018) (http://blanco.biomol.uci.edu/mpstruc/) (White, 2009). These correspond mainly to crystal structures and less frequently to Nuclear Magnetic Resonance (NMR) or cryogenic electron microscopy (cry-EM) structures. We discarded all non-transmembrane, monomeric and monotopic (not embedded in the lipid bilayer) proteins. Pre-processing of the dataset was performed by excluding single chains, dimers in which one of the chains was a soluble protein, single MPs interacting with soluble small peptides (protein-peptide), pores, protein-antibodies (since antibodies are soluble proteins) and proteins with small organic or non-organic ligands (protein-ligand). Additionally, were also excluded structures with unknown residues or with a high content in incomplete amino-acids, as well as structures with interfaces interacting highly with lipids or with an unclear PPI. Sequences were filtered to ensure at most 35% sequence redundancy in each interface by using the PISCES web-server to prevent repeated complexes (Wang and Dunbrack, 2003). The final dataset was composed of 63% homo-dimers and 37% heterodimers. From PDB files, all possible dimer combinations were extracted for the structures in which the number of chains was higher than two. The final dataset is composed of 201 protein dimer combinations (**Table S1**). The selected structures were then subjected to further preparation. In particular we: i) identified and removed residues outside the transmembrane domain; ii) removed unnecessary heteroatoms; iii) reversed mutated non-standard amino-acids (e.g. selenomethionine were mutated to methionine); iv) modelled incomplete structures (for structures in which the amount of amino-acids represented solely by the α–carbon was negligible, we performed homology modelling with Modeller (Webb and Sali, 2017 to build the full backbone and side-chains, taking the known structure as template and using the sequence); and v) added hydrogens to the structures. For these we used in house PyMOL (DeLano, 2015) and Visual Molecular Dynamics (VMD) scripts (Humphrey et al., 1996).

### Definition of interfacial and non-interfacial residues

The relative solvent accessibility (RSA) defined as the ratio between an amino-acid ASA value and its corresponding area in a Gly-X-Gly peptide was calculated using an in-house pipeline. Residues above a 0.20 RSA cut-off were considered as surface residues (Lins et al., 2003). We obtained 56.565 possible surface residues from a total of 94.239. Secondly, we considered as interfacial residues those for which the pairwise distance between any atom of chain A and any atom of chain B was below 5 Å, splitting surface residues into two classes: interfacial (15.726 residues) and non-interfacial ones (40.839 residues).

**Determination of sequence and structural features of all residues**

Evolutionary conservation of all sites was calculated using the Jensen-Shannon divergence (JSD) measure (Lin, 1991) of the Position-Specific Scoring Matrix (PSSM), which itself was calculated with a local deployment of PSI-BLAST (Altschul et al., 1997). JSD, **Equation 1,** compares the amino-acid distribution observed in PSSM matrix $p_{ia}$ with a background distribution $f_a$, in this case BLOSUM62.

**Equation 1.** Calculation of the JSD values.

$$JSD = H\left(\frac{p_{ia} + p_a}{2}\right) - \frac{1}{2}H(p_{ia}) - \frac{1}{2}H(f_a)$$

H( ) denotes the entropy of amino-acid distribution. The code provided by Capra *et al.* was introduced into the pipeline due to its high performance in comparison with other methods (Capra and Singh, 2007). This metric works on the premise that the highest similarly at each position along the PPSM profile corresponds to the highest JSD value and therefore to a more conserved residue (Urano et al., 2016).

The DSSP (Database of Secondary Structure assignments for all Proteins entries) (Touw et al., 2015) was used to calculate ASA of each amino-acid, "i", under complexed ($_{comp}ASA_i$) and monomeric ($_{mon}ASA_i$) forms. These values were also used to calculate $\Delta ASA_i$ **(Equation 2).**

**Equation 2.** Calculation of the ASA variation.

$$\Delta ASA_i = \left|_{comp}ASA_i - _{mon}ASA_i\right|$$

For further clarification, we also listed all $_{rel}ASA_i$ values **(Equation 3),** which allows the differentiation of residues with equal $\Delta ASA_i$ but with different absolute monomer ASA values (Martins et al., 2014; Melo et al., 2016; Munteanu et al., 2015).

**Equation 3.** Calculation of the $_{rel}ASA_i$ values.

$$_{rel}ASA_i = \frac{\Delta ASA_i}{_{mon}ASA_i}$$

The $_{comp}ASA_i$, $_{mon}ASA_i$ and $\Delta ASA_i$ values were then multiply by Sander and Rost amino-acid constants (ALA: 106, ARG: 248, ASN: 157, ASP: 163, CYS:135, GLN: 198, GLU: 194, GLY: 84, HIS: 184, ILE: 169, LEU: 165, LYS: 205, MET: 188, PHE: 197, PRO: 136, SER: 130, THR: 142, TRP: 227, TYR: 222, VAL: 142) (Rost and Sander, 1994).

To evaluate the mobility and therefore the stability of each residue, we extracted their temperature factors (*B*-factor) value from the PDB file of the analyzed structures (obtained directly from MPSTRUC) by deploying Biopython (Cock et al., 2009). Additionally, in order to understand the structural micro-environment of protein residues, we calculated the environmental *B*-factor using for each residue the average between its own *B*-factor and its neighbors in sequence considering +5 and -5 residues (a sliding window).

**Determination of structural descriptors of membrane protein-protein interface**

Close, hydrophobic and hydrogen contacts, salt-bridges and π-interactions were described using BINANA – Binding Analyzer, a Python-implemented algorithm that characterizes protein complexes (Durrant and McCammon, 2011). Close Contacts correspond to the number of pairs of atoms formed within 2.5 and 4.0 Å distance radius.

**Data normalization**

Since the composition of the dataset was not equally distributed across the three classes of MPs presented here, we defined a correction factor ($C_{factor}$), **Equation 4**, based on the concept of propensity score calculation, as presented by (Huang, 2014). This factor is defined as the ratio between the frequency of occurrence of residue $i$ in each one of the classes ($fi_{CLAS}$) and the frequency of occurrence of the total number of amino-acids in that class ($fi_{TOT}$). The obtained MP class-specific $C_{factor}$ were used to correct the various metrics described in the **Results** section by multiplying them by their respective $C_{factor}$ with the exception of $_{rel}$ASA.

**Equation 4.** Calculation of the correction factor.

$$C_{factor} = \frac{fi_{CLAS}}{fi_{TOT}}$$

**Web interface and statistics**

MENSAdb is a rich data visualization web application built using Python's *Flask*-based *Dash* visualization framework (by *Plotly*). The application enables users to explore a membrane-dimer dataset and provides graphical and tabular data formats that can be visually explored (by filtering, zooming, panning, etc.) and downloaded (both graphical and raw data). MENSAdb's real-time query features are supported by a MongoDB backend, which enables the application to query, filter and aggregate the dataset in multiple meaningful ways. To boost performance, a *Flask* caching layer is applied to support the complex queries required for visualization. To further ensure performance and security and support high-availability scenarios, all HTTP traffic directed at MENSAdb is served by the NGINX high performance web-server and load-balancer, which then routes it to multiple MENSAdb application instances.

For all plots, residues are ordered by increasing hydrophobicity based on the Kyte and Doolittle hydropathy index (Kyte et al., 1982). Descriptive statistics such as three quartiles (Q1, Q2 and Q3), average and standard deviation were obtained using Pandas, a Python library (McKinney, 2010). All the reported p-values were calculated through SciPy (https://docs.scipy.org/) using the T-test for the means of two independent samples (*ttest_ind*) functions. Further statistics were calculated for amino-acids sets split according to the hydrophilic and hydrophobic potential as: (a) charged – Asp, Glu, Lys, Arg; (b) positively charged – Lys, and Arg; (c) negatively charged – Asp and Glu; (d) polar – Ser, Thr, Asn, Gln, Tyr and His; (e) non-polar – Ala, Val, Ile, Leu, Met, Phe and Trp; aromatic – Phe, Trp, Tyr, His. Cys, Gly and Pro were not included in those subsets.

**RESULTS**

Considering the importance that Artificial Intelligence (AI) and Big Data are taking in the real world, there is a need for platforms, digital infrastructures, particularly cloud-based platforms, enabling researchers to collect, access and analyze interdisciplinary data. In this paper we describe a real time web-application that summarizes essential evolutionary and physico-chemical properties of membrane complexes to understand the basic principles underlying their formation.

**Membrane proteins composition**

The overall residue distribution in **Figure 2** shows that MPs have a higher content of hydrophobic and aromatic residues, such as leucine, alanine, valine, glycine, isoleucine and phenylalanine that account for 55% of all detected residues. This high hydrophobic content was also previously reported in several studies (Eilers et al., 2002; Saidijam et al., 2018; Ulmschneider and Sansom, 2001).

The overall distribution of individual residues of membrane proteins by amino acid type, **Figure 2.A,B1**, shows that GAS residues (Glycine, Alanine, Serine) (Zhang et al., 2015) are particularly enriched at the MPs non-surface. These small residues are the strong driving force for membrane folding (Zhang et al., 2009). As expected, charged residue are excluded from the MPs non-surface. The propensities for charged and polar residues at interfaces are intermediate between those for non-surface and non-interface surfaces. Residues distribution are in close agreement with several studies demonstrating that PPIs are mostly hydrophobic (e.g., leucine, isoleucine) in nature, with some aromatic residues (phenylalanine and tyrosine) and yield a buried non-polar surface area (Duarte et al., 2013; Ulmschneider and Sansom, 2001; Yan et al., 2008).

Evolutionary conservation of protein sequences is a key feature for understanding what are the functionally and structurally important residues in protein-protein interfaces. We used JSD dissimilarity score, in which values close to 0 mean a similar distribution whereas scores of 1 corresponds to totally discordant distributions. **Figure 2.B2** reveals that the highest JSD normalized values differences are for the more conserved GAS in the non-surface, and the non-polar residues in the interface. Additional results are available in "Conservation" option in the MENSAdb web-server.

Illustrative plots of both average *B*-factor values (by residue and using a five-residue window) can be found in the "Average *B*-factor" and "Environmental *B*-factor" options in the MENSAdb web-server. These factors measure the fluctuation of an atom around its mean position. Various authors have suggested that for soluble PPIs lower *B*-factors values for interfacial residues are indicative of lower flexibility (Chakravarty et al., 2015; Jones and Thornton, 1995; Liu et al., 2010). We observed a decrease in normalized *B*-factor values of the interfacial residues compared to the non-interfacial surface ones ($4.13\pm3.30$ vs $4.82\pm3.71$; p-value=$1.299*10^{-100}$), putting their average closer to the non-surface MP residues ($3.75\pm2.89$). Also, interfacial surface and non-surface positively charged residues are the most dissimilar ($2.91\pm2.10$ vs $1.03\pm0.79$; p-value=$1.149*10^{-13}$). The same holds true for environmental *B*-factor. These observations agree with findings attained for soluble PPIs (Chakravarty et al., 2015).

The ASA descriptors detect protein regions that, when interacting or aggregating, lose solvent accessible area. MENSAdb and **Figure 2.B3** shows that $_{rel}$ASA, which is the fractions of $\Delta$ASA by $_{mon}$ASA, is increased upon complex formation compared to the non-interfacial surface ($14.00\pm28.43$ vs $9.80 \pm 24.63$; p-value=$6.363*10^{-66}$), which seems particular relevant for charged residues. Additional and detailed information about $_{mon}$ASA, $_{comp}$ASA, $\Delta$ASA and $_{rel}$ASA can be view in MENSAdb web-server.

**Characteristics of interfacial residues**

Identification and characterization of critical features of membrane dimers PPIs can provide important clues to pinpoint particular residues or interactions, important for drug development. For this, additional interfacial structural characteristics were quantified to better understand MP dimers. Concerning the intermolecular atomic contacts per amino-acid type, we observed that the aromatic residues (normalized contacts at 4 Å: 0.47±0.51) are much more prone to establish close contacts at short distance than other residues. Arg was also highlighted in our results (normalized contacts at 4 Å: 0.76±0.83). For further information, check the "Interactions at 2.5 Angstroms" and "Interactions at 4.0 Angstroms" options in the MENSAdb web-server.

Hydrophobicity involving essentially large aromatic residues is key in MP dimers. In particular, Phe and Tyr establish π-π, t-stacking and cation-π in the different dimers. Cation-π interactions are also particularly relevant for Arg (for a closer detailed view, please see the "Hydrophobic Interactions", "Pi-Pi Interactions", "T-Stacking Interactions" and "Cation-Pi Interactions" options in the MENSAdb).

Additionally, although MPs residues reside in an apolar (low dielectric) environment (Lomize et al., 2007; Zhang et al., 2011), both salt-bridges between charged residues and hydrogen-bonds through almost all amino-acids are common to stabilize the interface and promote complex formation. Hydrogen-bonds measured here involving both side-chains and backbone are particularly important for polar (normalized according to Equation 4: 0.013±0.027) and charged residues (normalized values: 0.011±0.021) but also for aromatic ones (normalized values: 0.009±0.022), in particular tyrosine (normalized values: 0.011±0.026) and tryptophan (normalized values: 0.007±0.015). For a closer detailed view, please see the "Salt-bridge Interactions" and "Hydrogen-bond Interactions" options in MENSAdb web-server).

## CONCLUSION

MENSAdb is an open platform that includes a custom frontend application that users can interact with, the database and data export software components. It is the first database reporting a comprehensive and thorough structural and physic-chemical analysis of a curated collection of membrane protein dimer structures. The information is displayed in real time, in a user-friendly, interactive format, allowing an in-depth characterization of membrane dimer interfaces.

## AUTHOR CONTRIBUTIONS

## FUNDING

Irina S. Moreira acknowledges support by the Fundação para a Ciência e a Tecnologia (FCT) Investigator programme - IF/00578/2014 (co-financed by European Social Fund and Programa Operacional Potencial Humano). This work was also financed by the European Regional Development Fund (ERDF), through the Centro 2020 Regional Operational Programme under project CENTRO-01-0145-FEDER-000008: BrainHealth 2020. We also acknowledge the grants POCI-01-0145-FEDER-031356 and PTDC/QUI-OUT/32243/2017 financed by national funds through the FCT/MCTES and co-financed by the European Regional Development Fund (ERDF), namely under the following frameworks: "Projetos de Desenvolvimento e Implementação de Infraestruturas de Investigação inseridas no RNIE"; "Programa Operacional Competitividade e Internacionalização – POCI", "Programa Operacional Centro2020", and/or State Budget. Alexandre M.J.J.

Bonvin and P. Koukos acknowledge funding from the Dutch Foundation for Scientific Research (NWO) (TOP-PUNT grant 718.015.001).

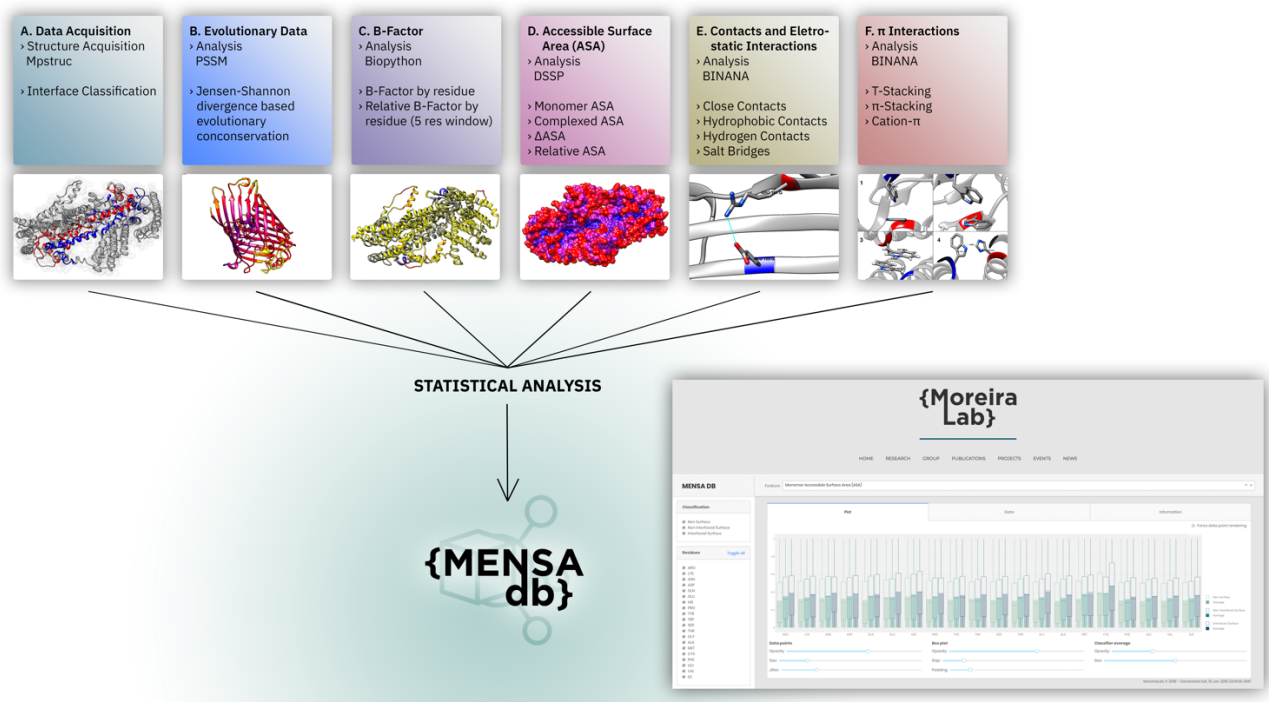

**Figure 1 - Overall representation of MENSAdb.** Boxes A through F illustrate the steps involving the data acquisition, evolutionary conservation, AS and PPI analysis. Each box is followed by an image of an example of the proteinic motifs under the scope of this work. (A) - interface between chains A and B of the STRA6 receptor for retinol uptake (PDBid: 5SY1) in *Danio rerio* (Chen et al., 2016). (B) - representation of evolutionary conservation of protein motifs (purple being more conserved and yellow less conserved) in the chain P of a hedgehog auto-processing domain in *Drosophila melanogaster* (PDBid: 1AT0) (Hall et al., 1997). (C) and (D) - complexed accessible surface area and average *B*-factor, respectively, of the chains A and B of 5SY1 (Chen et al., 2016). (E) – salt-bridge between GLU120 and ARG161 of the chain Q of the sucrose-specific porin (PDBid: 1A0T) of *Salmonella* Typhimurium (Forst et al., 1998). (F) - spectrum of π systems predicted. 1. and 2. T-stacking motif between TRP25 (chain L) and TRP255 (chain M) from *Rattus norvegicus* S100B protein (PDBid: 1XYD) (Wilder et al., 2005) is represented from two perspectives; 3. illustration of a π-π stacking structure between TRP262 (chain A) and TRP262 (chain B) from *Archaeoglobus fulgidus* CDP-alcohol phosphotransferase (PDBid: 4O6M) (Sciara et al., 2014); 4. cation-π interaction between HIS275 (chain B) and TRP175 (chain C) from *Escherichia coli* formate dehydrogenase-N (PDNid: 1KQF) (Jormakka et al., 2002). An overall depiction of the web platform presented in this work is also represented in the right bottom of the figure.

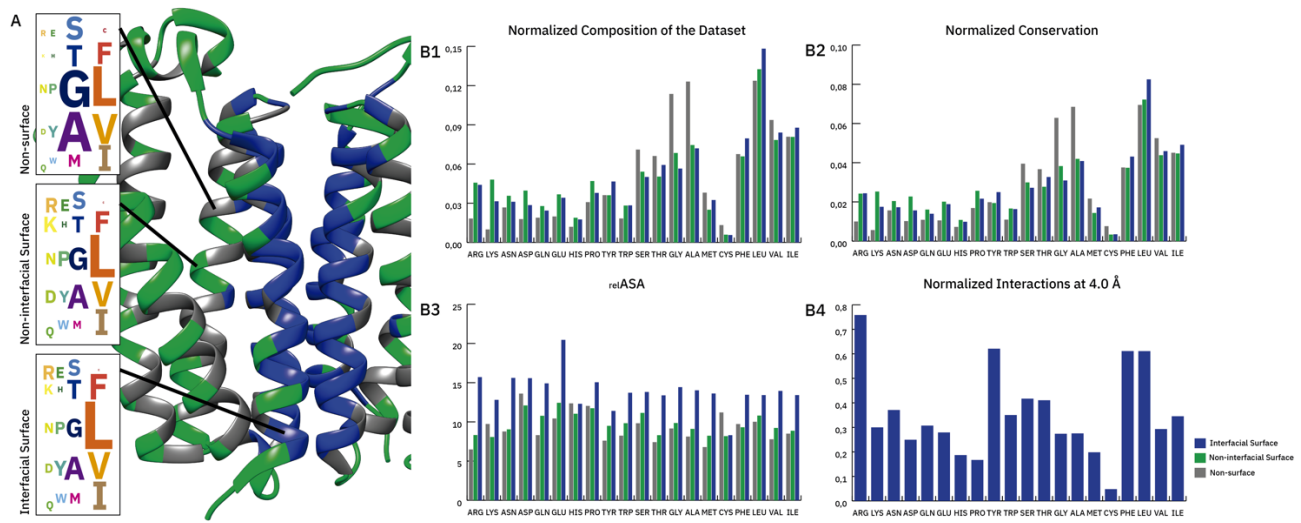

**Figure 2 – Panel of selected structural and physic-chemical properties of MPs and their interactions.** (A) – residue distribution of the translocator membrane protein (PDBid: 4UC1) from *Rhodobacter sphaeroides* (Li et al., 2015) . (B1) – normalized residue composition of the dataset. (B2) – normalized evolutionary conservation scores. (B3) –$_{rel}$ASA. (B4) – normalized close interactions at 4 Å.

**CONFLICT OF INTEREST**

The authors declare that the research was conducted in the absence of any commercial or financial relationships that could be construed as a potential conflict of interest.

**Table SI1.** Full membrane dimer dataset used in MENSAdb, discriminate by PDB code.

| PDB CODE | NUMBER OF DIMERS | DIMER CHAINS | CLASS | NAME | SPECIES | TAXONOMIC DOMAIN | RESOLUTION | UNIQUE CHAINS | SUBGROUP NAME | COMPLEX CLASS | OLIGOMER STATE | STOICHIOMETRY |
|---|---|---|---|---|---|---|---|---|---|---|---|---|
| 1A0T | 1 | PQ | beta | ScrY sucrose-specific porin | Salmonella typhimurium | Bacteria | 2.40 | 1 | Beta-Barrel Membrane Proteins: Porins and Relatives | protein-ligand | multimer | ['BA1: Homo 3-mer - A3', 'Asymmetry: C3'] |
| 1BL8 | 1 | AB | alpha | KcsA Potassium channel, H⁺ gated | Streptomyces lividans | Bacteria | 3.20 | 1 | Channels: Potassium, Sodium, & Proton Ion-Selective | protein-protein | multimer | ['Asymmetry: C4', 'BA1: Homo 4-mer - A4'] |
| 1EK9 | 1 | AB | beta | TolC outer membrane protein | Escherichia coli | Bacteria | 2.10 | 1 | Beta-Barrel Membrane Proteins: Monomeric/Dimeric | protein-protein | multimer | ['BA1: Homo 3-mer - A3', 'Asymmetry: C3'] |
| 1EYS | 2 | HM, LM | alpha | Photosynthetic Reaction Center | Thermochromatium tepidum | Bacteria | 2.20 | 4 | Photosynthetic Reaction Centers | protein-protein | multimer | ['Asymmetry: C1', 'BA1: Hetero 4-mer - ABCD'] |
| 1EZV | 2 | CD, CG | alpha | Cytochrome bc | Saccharomyces cegy5brevisiae | Eukaryota | 2.30 | 13 | Electron Transport Chain Complexes: Complex III | protein-antibody | multimer* | ['Asymmetry: C1', 'BA1: Hetero 20-mer - A2B2C2D2E2F2G2H2I2JK'] |
| 1KF6 | 1 | CD | alpha | E. coli Quinol-Fumarate Reductase with Bound Inhibitor HQNO | Escherichia coli (strain K12) | Bacteria | 2.70 | 8 | Oxidoreductases | protein-protein | multimer | ['Asymmetry: C1', 'BA1: Hetero 4-mer - ABCD', 'BA2: Hetero 4-mer - ABCD', 'BA3: Hetero 8-mer - A2B2C2D2'] |
| 1L7V | 1 | AB | alpha | BtuCD Vitamin B Transporter | Escherichia coli | Bacteria | 3.20 | 2 | ATP Binding Cassette (ABC) Transporters | protein-protein | multimer* | ['Asymmetry: A2B2', 'BA1: Hetero 4-mer - A2B2'] |
| 1LNQ | 1 | AB | alpha | MthK Potassium channel, Ca ⁺⁺ gated | Methanothermobacter thermautotrophicus | Archaea | 3.30 | 1 | Channels: Potassium, Sodium, & Proton Ion-Selective | protein-protein | multimer | ['Asymmetry: A8', 'BA1: Homo 8-mer - A8'] |
| 1MAL | 1 | AB | beta | LamB Maltoporin | Escherichia coli | Bacteria | 3.10 | 1 | Beta-Barrel Membrane Proteins: Porins and Relatives | protein-protein | multimer | ['BA1: Homo 3-mer - A3', 'Asymmetry: A3'] |
| 1NEK | 1 | CD | alpha | Succinate: quinone oxidoreductase (SQR, Complex II) | Escherichia coli | Bacteria | 2.60 | 4 | Electron Transport Chain Complexes: Complex II | protein-protein | multimer* | ['Asymmetry: BA1', 'BA1: Hetero 4-mer - ABCD', 'BA2: Hetero 12-mer - A3B3C3D3', 'BA3: Hetero 8-mer - A2B2C2D2'] |
| 1OCC | 10 | AB, AC, AE, AL, AM, BI, | alpha | Cytochrome C Oxidase | Bos taurus (bovine) heart mitochndria | Eukaryota | 2.80 | 13 | Electron Transport Chain Complexes: Complex IV | protein-protein | multimer | ['Asymmetry: A2B2C2D2E2F2G2H2I2J2K2L2M2', 'BA1: Hetero 26-mer - |

| | | CG, CJ, DK, LM | | | | | | | (Cytochrome C Oxidase) | | | A2B2C2D2E2F2G2H2I2J2 K2L2M2' |
|---|---|---|---|---|---|---|---|---|---|---|---|---|
| **1OTS** | 1 | AB | alpha | H⁺/Cl⁻ Exchange Transporter | Escherichia coli | Bacteria | 2.51 | 5 | Hcl Exchange Transporters | protein-antibody | multimer* | ['Asymmetry: A2B2C2', 'BA1: Hetero 6-mer - A2B2C2'] |
| **1P7B** | 1 | AB | alpha | KirBac1.1 Inward-Rectifier Potassium channel (closed state) | Burkholderia pseudomallei | Bacteria | 3.65 | 1 | Channels: Potassium, Sodium, & Proton Ion-Selective | protein-protein | m-m | ['Asymmetry: Homo 4-mer - A4'] |
| **1PRC** | 1 | LM | alpha | Photosynthetic Reaction Center | Blastochloris viridis | Bacteria | 2.30 | 4 | Photosynthetic Reaction Centers | protein-protein | multimer | ['Asymmetry: ABCD', 'BA1: Hetero 4-mer - ABCD', 'BA2: A2B2C2D2'] |
| **1RHZ** | 2 | AB, AC | alpha | SecYE β - translocon | Methanococcus jannaschii | Archaea | 3.50 | 3 | Sec and Translocase Proteins | protein-peptide | both | ['Asymmetry: ABC', 'BA1: Hetero 3-mer - ABC'] |
| **1UUN** | 1 | AB | beta | MspA mycobacterial porin | Mycobacterium smegmatis | Bacteria | 2.50 | 1 | Beta-Barrel Membrane Proteins: Porins and Relatives | protein-protein | m-m | ['BA1: Homo 8-mer - A8', 'Asymmetry: A2'] |
| **1XFH** | 1 | AB | alpha | Glutamate Transporter Homologue (Glt) | Pyrococcus horikoshii | Archaea | 3.50 | 1 | Amino Acid Secondary Transporters | protein-protein | multimer | ['Asymmetry: A3', 'BA1: Homo 3-mer - A3'] |
| **1YEW** | 2 | AB, BC | alpha | Particulate methane monooxygenase (pMMO) | Methylococcus capsulatus | Bacteria | 2.80 | 3 | Oxygenases | protein-protein | multimer | ['Asymmetry:  C1', 'BA1: Hetero 3-mer - ABC', 'BA2: Hetero 3-mer - ABC', 'BA3: Hetero 3-mer - ABC'] |
| **1ZLL** | 1 | AB | alpha | Phospholamban homopentamer | Homo sapiens | Eukaryota | NMR Structure | 2 | P-type ATPase | protein-protein | multimer | ['Asymmetry: C5', 'BA1: Homo 5-mer - A5'] |
| **1ZOY** | 1 | CD | alpha | Succinate:ubiquinone oxidoreductase (SQR, Complex II; pig heart) | Sus scrofa | Eukaryota | 2.40 | 4 | Electron Transport Chain Complexes: Complex II | protein-protein | multimer* | ['Asymmetry: ABCD', 'BA1: Hetero 4-mer - ABCD'] |
| **1ZRT** | 2 | CD, CP | alpha | Cytochrome bc | Rhodobacter capsulatus | Bacteria | 3.50 | 3 | Electron Transport Chain Complexes: Complex III | protein-protein | multimer | ['Asymmetry: A2B2C2', 'BA1: Hetero 6-mer - A2B2C2'] |
| **2AHY** | 1 | AB | alpha | NaK channel (Na⁺ complex) | Bacillus cereus | Bacteria | 2.40 | 1 | Channels: Potassium, Sodium, & Proton Ion-Selective | protein-protein | m-m | ['Asymmetry: Homo 4-mer - A4'] |
| **2BL2** | 1 | AB | alpha | Rotor of V-type Na⁺ -ATPase | Enterococcus hirae | Bacteria | 2.10 | 1 | Bacterial V-type ATPase | protein-protein | multimer | ['Asymmetry: Homo 10-mer - A10'] |
| **2BS2** | 1 | CF | alpha | Fumarate Reductase Complex | Wolinella succinogenes | Bacteria | 1.78 | 3 | Electron Transport Chain Complexes: Complex II | protein-protein | multimer | ['Asymmetry: A2B2C2', 'BA1: Hetero 6-mer - A2B2C2'] |

| 2FBW | 1 | CD | alpha | Succinate: ubiquinone oxidoreductase (SQR, Complex II; chicken heart) w. carboxin inhibitor | Gallus gallus | Eukaryota | 2.10 | 4 | Electron Transport Chain Complexes: Complex II | protein-protein | multimer* | ['Asymmetry: C1', 'BA1: Hetero 4-mer – ABCD', 'BA2: Hetero 4-mer – ABCD'] |
|------|---|----|-------|------|------|------|------|---|------|------|------|------|
| 2FYN | 2 | AB, AD | alpha | Cytochrome bc | Rhodobacter sphaeroides | Bacteria | 3.20 | 3 | Electron Transport Chain Complexes: Complex III | protein-protein | multimer | ['Asymmetry: A6B6C6', 'BA1: Hetero 6-mer - A2B2C2', 'BA2: A2B2C2', 'BA3: A2B2C2'] |
| 2GR8 | 1 | AC | beta | Trimeric autotransporter | Haemophilus influenzae | Bacteria | 2.00 | 1 | Outer Membrane Autotransporters | protein-protein | multimer | ['BA1: Homo 3-mer - A3', '2': 'A3', 'Asymmetry: A6'] |
| 2HAC | 1 | AB | alpha | Transmembrane dimer of the TCR-CD3 complex | Homo sapiens | Eukaryota | NMR Structure | 3 | Immune Receptors | protein-protein | m-m | ['Asymmetry: C2', 'BA1: Homo 2-mer – A2'] |
| 2HYD | 1 | AB | alpha | Sav1866 Multidrug Transporter | Staphylococcus aureus | Bacteria | 3.00 | 1 | ATP Binding Cassette (ABC) Transporters | protein-ligand | m-m | ['Asymmetry: A2', 'BA1: Homo 2-mer - A2'] |
| 2J1N | 1 | AB | beta | OmpC Osmoporin | Escherichia coli | Bacteria | 2.00 | 1 | Beta-Barrel Membrane Proteins: Porins and Relatives | protein-protein | multimer | ['BA1: Homo 3-mer - A3', 'Asymmetry: A3'] |
| 2J8S | 1 | AB | alpha | Drug Export Pathway of Multidrug Exporter AcrB Revealed by DARPin Inhibitors | Escherichia coli | Bacteria | 2.54 | 3 | Membrane protein: exporter | protein-protein | m-m | ['Asymmetry: Hetero 5-mer - A3B2','BA1: Hetero 5-mer - A3B2'] |
| 2K9Y | 1 | AB | alpha | EphA2 transmembrane segment dimer | Homo sapiens | Eukaryota | NMR Structure | 1 | Erythropoietin-Producing Hepatocellular Receptors | protein-protein | m-m | ['Asymmetry: Homo 2-mer - A2'] |
| 2KIX | 1 | AB | alpha | M2 proton channel (BM2) | Influenza B | Viruses | NMR Structure | 1 | Channels: Other Ion Channels | protein-protein | multimer | ['Asymmetry: C4', 'BA1: Homo 4-mer - A4'] |
| 2L35 | 1 | AB | alpha | DAP12 dimeric signaling domain in complex with activating receptor NKG2C | Homo sapiens | Eukaryota | NMR Structure | 2 | Immune Receptors | protein-protein | multimer | ['Asymmetry: C1', 'BA1: Hetero 2-mer – AB'] |
| 2LZL | 1 | AB | alpha | FGFR3 Fibroblast growth factor receptor 3 | Homo sapiens | Eukaryota | NMR Structure | 1 | Fibroblast Growth Factor Receptors | protein-protein | m-m | ['Asymmetry: Homo 2-mer - A2'] |

| | | | | transmembrane dimer | | | | | | | | |
|---|---|---|---|---|---|---|---|---|---|---|---|---|
| 2M59 | 1 | AB | alpha | VEGFR2 vascular endothelial growth factor receptor 2 transmembrane dimer | Homo sapiens | Eukaryota | NMR Structure | 1 | Vascular Endothelial Growth Factor Receptors | protein-protein | m-m | ['Asymmetry: Homo 2-mer - A2'] |
| 2MPN | 1 | AB | alpha | YgaP rhodanese homodimeric transmembrane domain | Escherichia coli | Bacteria | NMR Structure | 1 | Rhodaneses | protein-protein | m-m | ['Asymmetry: Homo 2-mer - A2'] |
| 2MPR | 1 | AB | beta | LamB Maltoporin | Salmonella typhimurium | Bacteria | 2.40 | 1 | Beta-Barrel Membrane Proteins: Porins and Relatives | protein-protein | multimer | ['BA1: Homo 3-mer - A3', 'Asymmetry: A3'] |
| 2NQ2 | 1 | AB | alpha | HI1470/1 Putative Metal-Chelate-type ABC Transporter | Haemophilus influenzae | Bacteria | 2.40 | 2 | ATP Binding Cassette (ABC) Transporters | protein-protein | multimer* | ['Asymmetry: A2B2', 'BA1: Hetero 4-mer - A2B2'] |
| 2NWL | 1 | AB | alpha | Aspartate Transporter Li+ - Bound State(Glt) | Pyrococcus horikoshii | Archaea | 2.96 | 1 | Amino Acid Secondary Transporters | protein-ligand | multimer | ['Asymmetry: A3', 'BA1: Homo 3-mer - A3'] |
| 2O4V | 1 | AB | beta | OprP phosphate-specific transporter | Pseudomonas aeruginosa | Bacteria | 1.90 | 1 | Beta-Barrel Membrane Proteins: Porins and Relatives | protein-ligand | multimer | ['BA1: Homo 3-mer - A3', 'Asymmetry: A3'] |
| 2ONK | 1 | CD | alpha | Molybdate Transporter ModB Complexed with ModA | Archaeoglobus fulgidus | Archaea | 3.10 | 3 | ATP Binding Cassette (ABC) Transporters | protein-protein | multimer* | ['Asymmetry: A4B4C2', 'BA1: Hetero 5-mer - A2B2C', 'BA2: A2B2C'] |
| 2PNO | 1 | AB | alpha | Leukotriene LTC Synthase in complex with glutathione | Homo sapiens | Eukaryota | 3.30 | 1 | Membrane-Associated Proteins in Eicosanoid and Glutathione Metabolism (MAPEG) | protein-ligand | multimer | ['Asymmetry: Homo 3-mer - A3', 'BA1: A3', 'BA2: A3', 'BA3: A3'] |
| 2Q7M | 1 | AB | alpha | 5-Lipoxygenase-Activating Protein (FLAP) with Bound MK-591 Inhibitor | Homo sapiens | Eukaryota | 4.00 | 1 | Membrane-Associated Proteins in Eicosanoid and Glutathione Metabolism (MAPEG) | protein-ligand | multimer | ['Asymmetry: A6', 'BA1: Homo 3-mer - A3', 'BA2: A3'] |
| 2R6G | 1 | FG | alpha | MalFGK -MBP Maltose uptake transporter complex | Escherichia coli | Bacteria | 2.80 | 4 | ATP Binding Cassette (ABC) Transporters | protein-protein | multimer* | ['Asymmetry: A2BCD', 'BA1: Hetero 5-mer - A2BCD'] |
| 2VL0 | 1 | AB | alpha | Prokaryotic pentameric ligand-gated ion channel (ELIC) | Erwinia chrysanthemi | Bacteria | 3.30 | 1 | Cys-Loop Receptor Family | protein-protein | multimer | ['Asymmetry: A10', 'BA1: Homo 5-mer - A5', 'BA2: A5'] |
| 2VPZ | 1 | CG | alpha | Polysulfide Reductase PsrABC (native) | Thermus thermophilus | Bacteria | 2.40 | 3 | MGD Oxidoreductases | protein-protein | multimer* | ['Asymmetry: A2B2C2', 'BA1: Hetero 6-mer - A2B2C2'] |
| 2WIE | 1 | AB | alpha | Rotor of H⁺ - dependent F-ATP Synthase of an alkaliphilic cyanobacterium | Spirulina platensis | Bacteria | 2.10 | 2 | F-type ATPase | protein-protein | multimer | ['Asymmetry: Homo 15-mer - A15'] |
| 2WIT | 1 | AB | alpha | BetP glycine betaine transporter | Corynebacterium glutamicum | Bacteria | 3.35 | 1 | Betaine/Choline/Carnitine Transporter (BCCT) Family | protein-ligand | multimer | ['Asymmetry: A3', 'BA1: Homo 3-mer - A3'] |
| 2WLJ | 1 | AB | alpha | KirBac3.1 Inward-Rectifier Potassium channel (semi-latched) | Magnetospirillum magnetotacticum | Bacteria | 2.60 | 1 | Channels: Potassium, Sodium, & Proton Ion-Selective | protein-protein | m-m* | ['Asymmetry: Homo 4-mer - A4'] |
| 2X2V | 1 | AB | alpha | Rotor of H⁺ - dependent F-ATP Synthase | Bacillus pseudofirmus OF4 | Bacteria | 2.50 | 1 | F-type ATPase | protein-protein | multimer | ['Asymmetry: A13', 'BA1: Homo 13-mer - A13'] |
| 2XND | 1 | OP | alpha | Fc-ring complex | Bos taurus | Eukaryota | 3.50 | 6 | F-type ATPase | protein-protein | multimer | ['Asymmetry: A8B3C3DEF', 'BA1: Hetero 17-mer - A8B3C3DEF'] |
| 2YEV | 2 | AB, AC | alpha | Cytochrome C Oxidase, | Thermus thermophilus | Bacteria | 2.36 | 3 | Electron Transport Chain Complexes: Complex IV (Cytochrome C Oxidase) | protein-protein | multimer* | ['Asymmetry: C1', 'BA1: Hetero 3-mer - ABC', 'BA2: Hetero 3-mer – ABC'] |
| 2ZT9 | 3 | AB, AG, CH | alpha | Cytochrome complex | Nostoc sp. PCC 7120 | Bacteria | 3.00 | 8 | Electron Transport Chain Complexes: Cytochrome b | protein-protein | multimer | ['Asymmetry: Hetero 16-mer - A4B2C2D2E2F2G2'] |
| 2ZW3 | 1 | AB | alpha | Connexin 26 (Cx26; GJB2) gap junction | Homo sapiens | Eukaryota | 3.50 | 1 | Channels: Gap Junctions | protein-protein | multimer | ['Asymmetry: A6', 'BA1: Homo 12-mer - A12'] |
| 2ZXE | 2 | AB, AG | alpha | Na,K-ATPase; shark | Squalus acanthias | Eukaryota | 2.40 | 3 | P-type ATPase | protein-protein | multimer* | ['Asymmetry: ABC', 'BA1: Hetero 3-mer - ABC'] |

| | | | | | | | | | | | |
|---|---|---|---|---|---|---|---|---|---|---|---|
| 2ZY9 | 1 | AB | alpha | Improved crystal structure of magnesium transporter MgtE | Escherichia coli | Bacteria | 2.94 | 2 | membrane protein metal transport | protein-protein | m-m | ['Asymmetry: Homo 2-mer - A2', 'BA1: Homo 2-mer - A2',] |
| 3A7K | 1 | AB | alpha | Halorhodopsin (HR) | Natronomonas pharaonis | Archaea | 2.00 | 1 | Bacterial and Algal Rhodopsins | protein-protein | multimer | ['Asymmetry: A3', 'BA1: Homo 3-mer - A3'] |
| 3B07 | 1 | AB | beta | &gamma -hemolysin composed of LukF and Hlg2 | Staphylococcus aureus | Bacteria | 2.50 | 2 | Adventitious Membrane Proteins: Beta-sheet Pore-forming Toxins/Attack Complexes | protein-protein | multimer | ['Asymmetry: A4B4', 'BA1: Hetero 8-mer - A4B4'] |
| 3B4R | 1 | AB | alpha | Site-2 Protease (S2P). Intramembrane Metalloprotease | Methanocaldococcus jannaschii | Archaea | 3.30 | 1 | Intramembrane Proteases | protein-protein | m-m* | ['Asymmetry: A2', 'BA1: Homo 2-mer - A2'] |
| 3B60 | 1 | AB | alpha | MsbA Lipid flippase with bound AMPPNP | Salmonella typhimurium | Bacteria | 3.70 | 1 | ATP Binding Cassette (ABC) Transporters | protein-protein | multimer | ['Asymmetry: Homo 2-mer - A2', 'BA1: A2'] |
| 3D31 | 1 | CD | alpha | ModBC Molybdate ABC Transporter in a trans-inhibited state | Methanosarcina acetivorans | Archaea | 3.00 | 2 | ATP Binding Cassette (ABC) Transporters | protein-protein | multimer* | ['Asymmetry: A2B2', 'BA1: Hetero 4-mer - A2B2'] |
| 3DH4 | 1 | AB | alpha | vSGLT Sodium Galactose Transporter | Vibrio parahaemolyticus | Bacteria | 2.70 | 1 | Solute Sodium Symporter (SSS) Family | protein-ligand | multimer | ['Asymmetry: Homo 4-mer - A4', 'BA1: A2', 'BA2: A2'] |
| 3DIN | 2 | CD, CE | alpha | SecYEG translocon in complex with SecA | Thermotoga maritima | Bacteria | 4.50 | 4 | Sec and Translocase Proteins | protein-protein | multimer | ['Asymmetry: C1', 'BA1: Hetero 4-mer – ABCD', 'BA2: Hetero 4-mer – ABCD', 'BA3: Hetero 4-mer – ABCD'] |
| 3KLY | 1 | AB | alpha | FocA formate transporter without formate | Vibrio cholerae | Bacteria | 2.10 | 1 | Channels : Formate/Nitrite Transporter (FNT) Family | protein-protein | multimer | ['Asymmetry: A5', 'BA1: Homo 5-mer - A5'] |
| 3MK7 | 2 | AB, AC | alpha | Cytochrome Oxidase | Pseudomonas stutzeri | Bacteria | 3.20 | 4 | Electron Transport Chain Complexes: Complex IV (Cytochrome C Oxidase) | protein-protein | multimer* | ['Asymmetry: C1', 'BA1: Hetero 4-mer – ABCD', 'BA2: Hetero 4-mer – ABCD'] |
| 3MP7 | 1 | AB | alpha | SecYE&beta translocon | Pyrococcus furiosus | Archaea | 3.10 | 2 | Sec and Translocase Proteins | protein-protein | m-m* | ['Asymmetry: AB', 'BA1: Hetero 2-mer - AB'] |
| 3ND0 | 1 | AB | alpha | H+ /Cl- Eukaryotic | Synechocystis sp. pcc 6803 | Bacteria | 3.20 | 1 | Hcl Exchange Transporters | protein-protein | m-m | ['Asymmetry: A2', 'BA1: Homo 2-mer - A2'] |

| | | | | Exchange Transporter | | | | | | | |
|---|---|---|---|---|---|---|---|---|---|---|---|
| **3OOR** | 1 | BC | alpha | Nitric Oxide Reductase | Pseudomonas aeruginosa | Bacteria | 2.70 | 6 | Nitric Oxide Reductases | protein-antibody | multimer* | ['Asymmetry: A2', 'BA1: Homo 2-mer - A2'] |
| **3O44** | 1 | AB | beta | Cytolysin pore-forming toxin | Vibrio cholerae | Bacteria | 2.88 | 1 | Adventitious Membrane Proteins: Beta-sheet Pore-forming Toxins/Attack Complexes | protein-protein | mltimer | ['Asymmetry: A14', 'BA1: Homo 7-mer - A7', 'BA2: A7'] |
| **3ODU** | 1 | AB | alpha | CXCR4 chemokine receptor complexed with IT1t antagonist | Homo sapiens | Eukaryota | 2.50 | 1 | G Protein-Coupled Receptors (GPCRs) | protein-ligand | m-m* | ['Asymmetry: A2', 'BA1: Homo 2-mer - A2'] |
| **3ORG** | 1 | AD | alpha | H⁺/Cl⁻ Eukaryotic Exchange Transporter | Cyanidioschyzon merolae | Eukaryota | 3.50 | 1 | Hcl Exchange Transporters | protein-protein | multimer | ['Asymmetry: A4', 'BA1: Homo 2-mer - A2', 'BA2: A2'] |
| **3P5N** | 1 | AB | alpha | RibU, S Component of the Riboflavin Transporter | Staphylococcus aureus | Bacteria | 3.60 | 1 | Energy-Coupling Factor (ECF) Transporters | protein-ligand | m-m | ['Asymmetry: A2', 'BA1: Homo 2-mer - A2'] |
| **3PJZ** | 1 | AB | alpha | TrkH potassium ion transporter | Vibrio parahaemolyticus | Bacteria | 3.51 | 1 | Superfamily of K Transporters (SKT proteins) | protein-protein | m-m | ['Asymmetry: A2', 'BA1: Homo 2-mer - A2'] |
| **3QF4** | 1 | AB | alpha | Heterodimeric ABC exporter TM287-TM288 | Thermotoga maritima | Bacteria | 2.90 | 2 | ATP Binding Cassette (ABC) Transporters | protein-ligand | m-m* | ['Asymmetry: C1', 'BA1: Hetero 2-mer AB'] |
| **3QNQ** | 1 | AB | alpha | ChbC EIIC phosphorylation-coupled saccharide transporter | Bacillus cereus | Bacteria | 3.30 | 1 | Phosphoenolpyruvate-Dependent Phosphotransferases (PTSs) | protein-ligand | multimer | ['Asymmetry: A4', 'BA1: Homo 2-mer - A2', 'BA2: A2'] |
| **3RHW** | 1 | AB | alpha | GlucClα anion-selective receptor (Fab-invermectin complex) | Caenorhabditis elegans | Eukaryota | 3.26 | 5 | Cys-Loop Receptor Family | protein-antibody | multimer* | ['Asymmetry: A5B5C5', 'BA1: Hetero 15-mer - A5B5C5'] |
| **3TDO** | 1 | AB | alpha | FNT3 Hydrosulphide Channel (HSC), pH 9.0 | Clostridium difficile | Bacteria | 2.20 | 1 | Channels : Formate/Nitrite Transporter (FNT) Family | protein-protein | multimer | ['Asymmetry: A5', 'BA1: Homo 5-mer - A5'] |
| **3TUI** | 1 | AB | alpha | Inward facing conformations of the MetNI methionine ABC transporter: | Escherichia coli (strain K12) | Bacteria | 2.90 | 4 | HYDROLASE/TRANSPORT PROTEIN | protein-protein | m-m | ['Asymmetry: Hetero 8-mer - A4B4', ' BA1: Hetero 4-mer - A2B2', 'BA2: Hetero 4-mer - A2B2'] |

| | | | | CY5 native crystal form | | | | | | | | |
|---|---|---|---|---|---|---|---|---|---|---|---|---|
| 3UKM | 1 | AB | alpha | Two-Pore Domain Potassium Channel (TWIK-1) | Homo sapiens | Eukaryota | 3.40 | 1 | Channels: Potassium, Sodium, & Proton Ion-Selective | protein-protein | multimer | ['Asymmetry: A4', 'BA1: Homo 2-mer - A2', 'BA2: A2'] |
| 3UX4 | 2 | AB, AC | alpha | Urel proton-gated inner membrane urea channel | Helicobacter pylori | Bacteria | 3.26 | 1 | Channels: Urea Transporters | protein-ligand | multimer | ['Asymmetry: Homo 6-mer - A6'] |
| 3VOU | 1 | AB | alpha | NaK channel chimera with grafted C-terminal region of a NaV channel | Bacillus weihenstephanensis (NaK) and Sulfitobacter pontiacus (NaV) | Bacteria | 3.20 | 1 | Channels: Potassium, Sodium, & Proton Ion-Selective | protein-protein | m-m | ['Asymmetry: Homo 4-mer - A4'] |
| 3VR8 | 1 | CD | alpha | Mitochondrial rhodoquinol-fumarate reductase | Ascaris suum | Eukaryota | 2.81 | 4 | Oxidoreductases | protein-protein | multimer* | ['Asymmetry: C1', 'BA1: Hetero 4-mer - ABCD', 'BA2: Hetero 4-mer - ABCD'] |
| 4A01 | 1 | AB | alpha | H⁺-translocating M-PPase | Vigna radiata | Eukaryota | 2.35 | 1 | Membrane-Integral Pyrophosphatases (M-PPases) | protein-ligand | m-m | ['Asymmetry: A2', 'BA1: Homo 2-mer - A2'] |
| 4AV3 | 1 | AB | alpha | Na⁺-translocating M-PPase with metal ions in active site | Thermotoga maritima | Bacteria | 2.60 | 1 | Membrane-Integral Pyrophosphatases (M-PPases) | protein-protein | m-m | ['Asymmetry: A2', 'BA1: Homo 2-mer - A2'] |
| 4COF | 1 | AB | alpha | GABAR receptor (3 homopentamer) | Homo sapiens | Eukaryota | 2.97 | 1 | Cys-Loop Receptor Family | protein-protein | multimer | ['Asymmetry: A5', 'BA1: Homo 5-mer - A5'] |
| 4CZB | 1 | AB | alpha | NhaP1 Na⁺/H⁺ antiporter, pH 8 | Methanocaldococcus jannaschii | Archaea | 3.50 | 2 | Antiporters | protein-protein | multimer | ['Asymmetry: A4', 'BA1: Homo 2-mer - A2', 'BA2: A2'] |
| 4DJH | 1 | AB | alpha | K - opioid receptor in complex with JDTic | Homo sapiens | Eukaryota | 2.90 | 1 | G Protein-Coupled Receptors (GPCRs) | protein-ligand | m-m* | ['Asymmetry: A2', 'BA1: Homo 2-mer - A2', 'BA2: A2'] |
| 4EV6 | 1 | AB | alpha | CorA Mg²⁺Transporter | Methanocaldococcus jannaschii | Archaea | 3.20 | 1 | CorA Superfamily Ion Transporters | protein-protein | multimer | ['Asymmetry: A5', 'BA1: Homo 5-mer - A5'] |
| 4EZC | 1 | AB | alpha | UT-B Urea Transporter | Bos taurus | Eukaryota | 2.36 | 1 | Channels: Urea Transporters | protein-ligand | multimer | ['Asymmetry: A3', 'BA1: Homo 3-mer - A3'] |
| 4F4L | 1 | AC | alpha | Voltage-Gated Sodium Channel (Na) | Magnetococcus marinus | Bacteria | 3.49 | 1 | Channels: Potassium, Sodium, & Proton Ion-Selective | protein-protein | multimer | ['Asymmetry: A4', 'BA1: Homo 4-mer - A4'] |

| | | | | | | | | | | | |
|---|---|---|---|---|---|---|---|---|---|---|---|
| 4F4S | 1 | AB | alpha | Structure of the yeast F1Fo ATPase c10 ring with bound oligomycin | Saccharomyces cerevisiae (strain ATCC 204508 / S288c) | Bacteria | 1.90 | 10 | MEMBRANE PROTEIN/ANTIBIOTIC | protein-protein | multimer | ['Asymmetry: Homo 10-mer - A10 ', 'BA1: Homo 10-mer - A10', 'BA2: Homo 10-mer - A10'] |
| 4G1U | 1 | AB | alpha | HmuUV heme transporter | Yersinia pestis | Bacteria | 3.00 | 2 | ATP Binding Cassette (ABC) Transporters | protein-protein | multimer* | ['Asymmetry: A2B2', 'BA1: Hetero 4-mer - A2B2'] |
| 4GX0 | 1 | AB | alpha | GsuK multi-ligand gated K+ channel, L97D mutant | Geobacter sulfurreducens | Bacteria | 2.60 | 1 | Channels: Potassium, Sodium, & Proton Ion-Selective | protein-protein | multimer | ['Asymmetry: Homo 4-mer - A4', 'BA1: A4'] |
| 4HKR | 1 | AB | alpha | Orai Calcium release-activated calcium (CRAC) channel | Drosophila melanogaster | Eukaryota | 3.35 | 1 | Channels: Calcium Ion-Selective | protein-protein | m-m | ['Asymmetry: Homo 6-mer - A6'] |
| 4HYG | 1 | AB | alpha | PSH presenilin/SPP homologue aspartate protease (C222 space group) | Methanoculleus marisnigri | Archaea | 3.32 | 1 | Intramembrane Proteases | protein-protein | multimer | ['Asymmetry: Homo 4-mer - A4', 'BA1: A4'] |
| 4J72 | 1 | AB | alpha | MraY phospho-MurNAc-pentapeptide translocase | Aquifex aeolicus | Bacteria | 3.30 | 1 | PNPT Superfamily | protein-protein | m-m | ['Asymmetry: A2', 'BA1: Homo 2-mer - A2'] |
| 4J7C | 1 | IJ | alpha | KtrAB potassium ion transporter | Bacillus subtilis | Bacteria | 3.50 | 2 | Superfamily of K Transporters (SKT proteins) | protein-protein | multimer* | ['Asymmetry: A8B4', 'BA1: Hetero 10-mer - A8B2', 'BA2: A8B2', 'BA3: A8B4'] |
| 4JKV | 1 | AB | alpha | Smoothened (SMO) receptor with bound antagonist, LY2940680 | Homo sapiens | Eukaryota | 2.45 | 1 | G Protein-Coupled Receptors (GPCRs) | protein-ligand | m-m* | ['Asymmetry: A2', 'BA1: Homo 2-mer - A2'] |
| 4JQ6 | 2 | AB, AC | alpha | Proteorhodopsin (blue-light absorbing), BPR | uncultured bacterium | Bacteria | 2.31 | 1 | Bacterial and Algal Rhodopsins | protein-protein | multimer | ['Asymmetry: Homo 6-mer - A6'] |
| 4KLY | 2 | AB, BC | alpha | Proteorhodopsin (blue-light absorbing); BPR, D97N mutant | gamma proteobacterium | Bacteria | 2.70 | 1 | Bacterial and Algal Rhodopsins | protein-protein | multimer | ['Asymmetry: A5', 'BA1: Homo 5-mer - A5'] |
| 4MBS | 1 | AB | alpha | CCR5 chemokine receptor with bound Maraviroc | Homo sapiens | Eukaryota | 2.71 | 1 | G Protein-Coupled Receptors (GPCRs) | protein-ligand | m-m | ['Asymmetry: A2', 'BA1: Monomer', 'BA2: A'] |

| | | | | | | | | | | | |
|---|---|---|---|---|---|---|---|---|---|---|---|
| **4MRN** | 1 | AB | alpha | Atm1-type ABC exporter, apo protein | Novosphingobium aromaticivorans | Bacteria | 2.50 | 1 | ATP Binding Cassette (ABC) Transporters | protein-protein | m-m | ['Asymmetry: A2', 'BA1: Homo 2-mer - A2'] |
| **4MT4** | 1 | AB | beta | CmeC bacterial multi-drug efflux transporter outer membrane channel | Campylobacter jejuni | Bacteria | 2.37 | 1 | Beta-Barrel Membrane Proteins: Monomeric/Dimeric | protein-protein | multimer | ['BA1: Homo 3-mer - A3', 'Asymmetry: A3'] |
| **4MYC** | 1 | AB | alpha | Atm1 mitochondrial ABC transporter, apo form | Saccharomyces cerevisiae | Eukaryota | 3.06 | 1 | ATP Binding Cassette (ABC) Transporters | protein-protein | multimer | ['Asymmetry: Homo 2-mer - A2', 'BA1: A2'] |
| **4O6M** | 1 | AB | alpha | AF2299 CDP-alcohol phosphotransfe rase w. bound CMP | Archaeoglobus fulgidus | Archaea | 1.90 | 1 | CDP-Alcohol Phosphotransferases | protein-ligand | m-m | ['Asymmetry: A2', 'BA1: Homo 2-mer - A2'] |
| **4O6Y** | 1 | AB | alpha | Cytochrome b | Arabidopsis thaliana | Eukaryota | 1.70 | 1 | Oxidoreductases | protein-ligand | m-m | ['Asymmetry: A2', 'BA1: Homo 2-mer - A2'] |
| **4P6V** | 6 | BD, BE, CD, CF, DE, EF | alpha | Na⁺ -pumping NADH:quinone oxidoreductase (Na⁺ -NQR) | Vibrio cholerae | Bacteria | 3.50 | 6 | Oxidoreductases | protein-protein | multimer | ['Asymmetry: ABCDEF', 'BA1: Hetero 6-mer - ABCDEF'] |
| **4PHZ** | 2 | AB, BC | alpha | Crystal structure of particulate methane monooxygenas e from Methylocystis sp. ATCC 49242 (Rockwell) | Methylococcus capsulatus | Bacteria | 2.59 | 12 | Oxidoreductases | protein-protein | multimer* | ['Asymmetry: Hetero 11-mer - A3B3C3D2', 'BA1: Hetero 11-mer - A3B3C3D2 '] |
| **4PIR** | 1 | AB | alpha | Serotonin 5-HT receptor | Mus musculus | Eukaryota | 3.50 | 2 | Cys-Loop Receptor Family | protein-protein | multimer | ['Asymmetry: A5B5', 'BA1: Hetero 10-mer - A5B5'] |
| **4PL0** | 1 | AB | alpha | McjD antimicrobial peptide transporter | Escherichia coli | Bacteria | 2.70 | 1 | ATP Binding Cassette (ABC) Transporters | protein-ligand | m-m | ['Asymmetry: A2', 'BA1: Homo 2-mer - A2'] |
| **4QNC** | 1 | AB | alpha | semiSWEET transporter in occluded state | Leptospira biflexa | Bacteria | 2.39 | 1 | SWEET and semiSWEET Transporters, and Their Relatives | protein-protein | m-m | ['Asymmetry: A2', 'BA1: Homo 2-mer - A2'] |
| **4QTN** | 1 | AB | alpha | PnuC vitamin B transporter | Neisseria mucosa | Bacteria | 2.80 | 1 | SWEET and semiSWEET | protein-ligand | multimer | ['Asymmetry: A3', 'BA1: Homo 3-mer - A3'] |

| | | | | | | | | | Transporters, and Their Relatives | | | |
|---|---|---|---|---|---|---|---|---|---|---|---|---|
| 4R0C | 1 | AB | alpha | YdaH transporter | Alcanivorax borkumensis | Bacteria | 2.96 | 1 | AbgT Family of Transporters | protein-protein | multimer | ['Asymmetry: A4', 'BA1: Homo 4-mer - A4'] |
| 4RDQ | 1 | AB | alpha | Bestrophin-1 (BEST1) Ca²⁺-activated Cl⁻ channel | Gallus gallus | Eukaryota | 2.85 | 5 | Channels: Other Ion Channels | protein-antibody | multimer* | ['Asymmetry: A5B5C5', 'BA1: Hetero 15-mer - A10B5'] |
| 4RI2 | 1 | AB | alpha | PsbS photoprotection protein | Spinacia oleracea | Eukaryota | 2.35 | 1 | Photoprotection Proteins | protein-ligand | m-m | ['Asymmetry: A2', 'BA1: Homo 2-mer - A2'] |
| 4RNG | 1 | AC | alpha | semiSWEET transporter in occluded state | Thermodesulfovibrio yellowstonii | Bacteria | 2.40 | 1 | SWEET and semiSWEET Transporters, and Their Relatives | protein-protein | multimer | ['Asymmetry: Homo 2-mer - A2', 'BA1: A2', 'BA2: A2', 'BA3: A6'] |
| 4RY2 | 1 | AB | alpha | Peptidase-containing ABC transporter (PCAT) | Ruminiclostridium thermocellum | Bacteria | 3.61 | 1 | ATP Binding Cassette (ABC) Transporters | protein-protein | m-m | ['Asymmetry: A2', 'BA1: Homo 2-mer - A2'] |
| 4TQU | 1 | MN | alpha | Alginate transporter AlgM1M2SS with bound periplasmic protein AlgQ2 | Sphingomonas sp. | Bacteria | 3.20 | 4 | ATP Binding Cassette (ABC) Transporters | protein-protein | multimer* | ['Asymmetry: A2BCD', 'BA1: Hetero 5-mer - A2BCD'] |
| 4UC1 | 1 | AB | alpha | Translocator protein (TSPO), A139T SeMet1 C121 | Rhodobacter sphaeroides | Bacteria | 1.80 | 1 | Translocator Protein (18 kDA) TSPO | protein-ligand | multimer | ['Asymmetry: Homo 2-mer - A2', 'BA1: A2'] |
| 4UV3 | 1 | AB | beta | CsgG bacterial amyloid secretion channel | Escherichia coli | Bacteria | 3.59 | 1 | Beta-Barrel Membrane Proteins: Monomeric/Dimeric | protein-protein | multimer | ['BA1: Homo 9-mer - A9', 'BA2:A9', 'Asymmetry: A18'] |
| 4WD7 | 1 | AB | alpha | KpBest Bestrophin homolog of the BEST1 Ca²⁺-activated Cl⁻ channel (ΔC7) | Klebsiella pneumoniae | Bacteria | 2.90 | 1 | Channels: Other Ion Channels | protein-protein | multimer | ['Asymmetry: A5', 'BA1: Homo 5-mer - A5'] |
| 4WFE | 1 | AB | alpha | Human TRAAK K+ channel in a K+ bound conductive conformation | Homo sapiens | Eukaryota | 2.50 | 6 | Metal transporter | protein-antibody | multimer* | ['Asymmetry: Hetero 6-mer - A2B2C2', 'BA1: Hetero 6-mer - A2B2C2'] |
| 4WGV | 1 | AC | alpha | SLC11 (NRAMP) transition-metal ion transporter in | Staphylococcus capitis | Bacteria | 3.10 | 2 | Solute Carrier (SLC) Transporter Superfamily | protein-antibody | multimer* | ['Asymmetry: A2B2', 'BA1: Hetero 4-mer - A2B2'] |

| | | | | complex with nanobodies | | | | | | | | |
|---|---|---|---|---|---|---|---|---|---|---|---|---|
| 4WIS | 1 | AB | alpha | TMEM16 Ca²⁺-activated lipid scramblase, crystal form 1 | Nectria haematococca | Eukaryota | 3.30 | 1 | TMEM16 Family Proteins | protein-protein | m-m | ['Asymmetry: A2', 'BA1: Homo 2-mer - A2'] |
| 4X5M | 1 | BC | alpha | semiSWEET transporter in inward-open conformation (crystal I) | Escherichia coli | Bacteria | 2.00 | 1 | SWEET and semiSWEET Transporters, and Their Relatives | protein-protein | multimer | ['Asymmetry: Homo 2-mer - A2', 'BA1: A2'] |
| 4XYD | 1 | AB | alpha | Nitric Oxide Reductase BC complex | Roseobacter denitrificans | Bacteria | 2.85 | 2 | Nitric Oxide Reductases | protein-ligand | m-m* | ['Asymmetry: AB', 'BA1: Hetero 2-mer - AB'] |
| 4YMS | 1 | CD | alpha | Art(QN) amino acid importer | Caldanaerobacter tengcongensis | Bacteria | 2.80 | 2 | ATP Binding Cassette (ABC) Transporters | protein-protein | multimer* | ['Asymmetry: A2B2', 'BA1: Hetero 4-mer - A2B2'] |
| 4YZF | 1 | AB | alpha | Erythrocyte Band 3 anion exchanger | Homo sapiens | Eukaryota | 3.50 | 3 | Solute Carrier Family 4 (anion exchanger) | protein-antibody | multimer* | ['Asymmetry: A4B4C4', 'BA1: Hetero 6-mer - A4B2', 'BA2: A2B2C2'] |
| 5A1S | 1 | AB | alpha | CitS Citrate symporter | Salmonella enterica | Bacteria | 2.50 | 1 | Solute Carrier (SLC) Transporter Superfamily | protein-peptide | multimer* | ['Asymmetry: A4', 'BA1: Homo 2-mer - A2', 'BA2: A2'] |
| 5A2N | 1 | AB | alpha | NRT1.1 nitrate transporter, apo form | Arabidopsis thaliana | Eukaryota | 3.70 | 1 | Major Facilitator Superfamily (MFS) Transporters | protein-protein | m-m | ['Asymmetry: A2', 'BA1: Homo 2-mer - A2'] |
| 5AEX | 1 | AB | alpha | Mep2 ammonium transceptor | Saccharomyces cerevisiae | Eukaryota | 3.20 | 1 | Channels: Amt/Mep/Rh proteins | protein-protein | multimer | ['Asymmetry: Homo 3-mer - A3', 'BA1: A3', 'BA2: A3'] |
| 5AWW | 2 | AB, AC | alpha | Precise Resting State of Thermus thermophilus SecYEG | Thermus thermophilus (strain HB8 / ATCC 27634 / DSM 579) | Bacteria | 2.72 | 3 | PROTEIN TRANSPORT/IMMUNE SYSTEM | protein-protein | multimer | ['Asymmetry: Hetero 3-mer - ABC', 'BA1: Hetero 3-mer - ABC'] |
| 5AZS | 1 | AB | beta | OprJ drug discharge outer membrane protein | Pseudomonas aeruginosa | Bacteria | 3.10 | 1 | Beta-Barrel Membrane Proteins: Monomeric/Dimeric | protein-protein | multimer | ['BA1: Homo 3-mer - A3', 'Asymmetry: A3'] |
| 5B57 | 1 | AB | alpha | BhuU/BhuV haem importer, inward facing | Burkholderia cenocepacia | Bacteria | 2.80 | 2 | ATP Binding Cassette (ABC) Transporters | protein-protein | multimer* | ['Asymmetry: A2B2', 'BA1: Hetero 4-mer - A2B2'] |
| 5BUN | 1 | AB | beta | ST50 discharge outer membrane protein | Salmonella enterica | Bacteria | 2.98 | 1 | Beta-Barrel Membrane Proteins: Monomeric/Dimeric | protein-protein | multimer | ['BA1: Homo 3-mer - A3', 'Asymmetry: A3'] |
| 5C78 | 1 | AD | alpha | PglK lipid-linked oligosaccharide flippase, apo- | Campylobacter jejuni | Bacteria | 2.90 | 1 | ATP Binding Cassette (ABC) Transporters | protein-protein | multimer | ['Asymmetry: A4', 'BA1: Homo 2-mer - A2', 'BA2: A2'] |

| | | | | | | | | | | | |
|---|---|---|---|---|---|---|---|---|---|---|---|
| | | | | inward structure 1 | | | | | | | |
| 5C8J | 1 | IL | alpha | DUF106 YidC-like protein | Methanocaldococcus jannaschi | Archaea | 3.50 | 5 | Sec and Translocase Proteins | protein-antibody | multimer* | ['Asymmetry: C1', 'BA1: Hetero 3-mer – ABC', 'BA2: Hetero 3-mer – ABC', 'BA3: Hetero 3-mer – ABC', 'BA4: Hetero 3-mer – ABC'] |
| 5CFB | 1 | AB | alpha | Human glycine receptor (hGlyR-α; 3) in complex with strychnine | Homo sapiens | Eukaryota | 3.04 | 1 | Cys-Loop Receptor Family | protein-protein | multimer | ['Asymmetry: A5', 'BA1: Homo 5-mer - A5'] |
| 5CTG | 1 | AB | alpha | SWEET transporter in a homotrimeric complex | Oryza sativa | Eukaryota | 3.10 | 1 | SWEET and semiSWEET Transporters, and Their Relatives | protein-protein | multimer | ['Asymmetry: A3', 'BA1: Homo 3-mer - A3'] |
| 5DO7 | 1 | AB | alpha | ABCG5/ABCG8 sterol transporter | Homo sapiens | Eukaryota | 3.93 | 2 | ATP Binding Cassette (ABC) Transporters | protein-protein | multimer* | ['Asymmetry: A2B2', 'BA1: Hetero 2-mer - AB', 'BA2: AB'] |
| 5EKP | 1 | AB | alpha | GtrB polyisoprenyl-glycosyltransferase (PI-GT) | Synechocystis sp. PCC6803 | Bacteria | 3.19 | 1 | Glycosyltransferases | protein-protein | multimer | ['Asymmetry: A4', 'BA1: Homo 4-mer - A4'] |
| 5EUL | 1 | EY | alpha | SecYE translocon in complex with SecA | Geobacillus thermodenitrificans | Bacteria | 3.70 | 4 | Sec and Translocase Proteins | protein-protein | multimer* | ['Asymmetry: ABCD', 'BA1: Hetero 4-mer - ABCD'] |
| 5H3O | 1 | AB | alpha | Cyclic-nucleotide-gated (CNG) channel | Caenorhabditis elegans | Eukaryota | 3.50 | 1 | Channels: Potassium, Sodium, & Proton Ion-Selective | protein-ligand | multimer | ['Asymmetry: A4', 'BA1: Homo 4-mer - A4'] |
| 5HK7 | 1 | AB | alpha | Bacterial sodium channel pore | Alkalilimnicola ehrlichii (strain ATCC BAA-1101 / DSM 17681 / MLHE-1) | Bacteria | 2.95 | 4 | Transport protein | protein-protein | multimer | ['Asymmetry: C4', 'BA1: Homo 4-mer - A4'] |
| 5J4I | 1 | AB | alpha | Crystal Structure of the L-arginine/agmatine antiporter from E. coli at 2.2 Angstroem resolution | Escherichia Coli | Bacteria | 2.21 | 2 | Transport protein | protein-protein | m-m | ['Asymmetry: Homo 2-mer - A2', 'BA1: Homo 2-mer - A2'] |
| 5KBN | 1 | AB | alpha | Fluc F- ion channel homolog in complex Ec2-S9 | Escherichia coli | Bacteria | 2.48 | 2 | Channels: Fluc Family | protein-protein | multimer* | ['Asymmetry: C1', 'BA1: Hetero 2-mer - AB', 'BA2: Hetero 2-mer - AB'] |

| | | | | | | | | | | | |
|---|---|---|---|---|---|---|---|---|---|---|---|
| | | | | monobody F801 mutant | | | | | | | |
| 5KHN | 1 | AB | alpha | HpnN hopanoid transporter (crystal form I) | Burkholderia multivorans | Bacteria | 3.44 | 1 | Multi-Drug Efflux Transporters | protein-protein | m-m | ['Asymmetry: A2', 'BA1: Homo 2-mer - A2'] |
| 5KXI | 2 | AB, AE | alpha | Nicotinic Acetylcholine α4β2 Receptor | Homo sapiens | Eukaryota | 3.94 | 2 | Cys-Loop Receptor Family | protein-ligand | multimer | ['Asymmetry: C1', 'BA1: Hetero 5-mer - A3B2'] |
| 5L22 | 1 | AB | alpha | PrtD Type-1 secretion system ABC transporter | Aquifex aeolicus | Bacteria | 3.15 | 1 | ATP Binding Cassette (ABC) Transporters | protein-ligand | m-m | ['Asymmetry: A2', 'BA1: Homo 2-mer - A2'] |
| 5MKK | 1 | AB | alpha | TmrAB antigen transporter homolog | Thermus thermophilus | Bacteria | 2.70 | 2 | ATP Binding Cassette (ABC) Transporters | protein-protein | m-m* | ['Asymmetry: C1', 'BA1: Homo 2-mer - AB'] |
| 5MRW | 3 | AB, AC, BD | alpha | Potassium-importing KdpFABC membrane complex | Escherichia coli | Bacteria | 2.90 | 12 | Superfamily of K Transporters (SKT proteins) | protein-protein | multimer* | ['Asymmetry: A3B3C3D3', 'BA1: Hetero 4-mer - ABCD', 'BA2: ABCD', 'BA3: ABCD'] |
| 5N77 | 1 | AB | alpha | CorA Mg²⁺ Transporter cytoplasmic domain with bound Mg²⁺ | Escherichia coli | Bacteria | 2.80 | 1 | CorA Superfamily Ion Transporters | protein-protein | multimer | ['Asymmetry: A5', 'BA1: Homo 5-mer - A5'] |
| 5NKQ | 1 | AB | alpha | Crystal structure of a dual topology fluoride ion channel. | Bordetella pertussis (strain Tohama I / ATCC BAA-589 / NCTC 13251) | Bacteria | 2.17 | 4 | Transport protein | protein-protein | multimer | ['Asymmetry: Hetero 8-mer - A4B4', 'BA1: Hetero 4-mer - A2B2', 'BA2: Hetero 4-mer - A2B2'] |
| 5SV0 | 1 | AB | alpha | ExbB/ExbD complex associated with TonB complex, pH 7.0 | Escherichia coli | Bacteria | 2.60 | 1 | Channels: Other Ion Channels | protein-protein | multimer | ['Asymmetry: A10', 'BA1: Homo 5-mer - A5', 'BA2: A5'] |
| 5SY1 | 1 | AB | alpha | STRA6 retinol-uptake receptor in complex with calmodulin (CaM) | Danio rerio | Eukaryota | 3.90 | 2 | Novel Receptors | protein-protein | multimer | ['Asymmetry: A2B2', 'BA1: Hetero 4-mer - A2B2'] |
| 5T0O | 1 | AB | alpha | CmeB multi-drug efflux transporter, C2 space group | Campylobacter jejuni | Bacteria | 3.15 | 1 | Multi-Drug Efflux Transporters | protein-protein | multimer | ['Asymmetry: A3', 'BA1: Homo 3-mer - A3'] |
| 5TIN | 1 | AB | alpha | Crystal Structure of Human Glycine Receptor alpha-3 Mutant N38Q | Homo sapiens | Eukaryota | 2.61 | 5 | Transport protein | protein-protein | multimer | ['Asymmetry: Homo 5-mer - A5', 'BA1: Homo 5-mer - A5'] |

| | | | | | | | | | | | |
|---|---|---|---|---|---|---|---|---|---|---|---|
| | | | | Bound to AM-3607 | | | | | | | |
| 5TQQ | 1 | AB | alpha | CLC-K chloride ion channel, class 1 | Bos taurus | Eukaryota | 3.76 | 5 | Channels: Other Ion Channels | protein-antibody | multimer* | ['Asymmetry: A2B2C2', 'BA1: Hetero 6-mer - A4B2'] |
| 5U1D | 1 | AB | alpha | Transporter associated with antigen processing (TAP) bound to ICP47 | Homo sapiens | Eukaryota | 4.00 | 3 | ATP Binding Cassette (ABC) Transporters | protein-peptide | both | ['Asymmetry: ABC', 'BA1: Hetero 3-mer - ABC'] |
| 5U6O | 1 | AB | alpha | HCN1 hyperpolarization-activated channel | Homo sapiens | Eukaryota | 3.50 | 1 | Channels: Potassium, Sodium, & Proton Ion-Selective | protein-protein | multimer | ['Asymmetry: A4', 'BA1: Homo 4-mer - A4'] |
| 5UNI | 1 | AB | alpha | Critical role of water molecules for proton translocation of the membrane-bound transhydrogenase | Thermus thermophilus (strain HB27 / ATCC BAA-163 / DSM 7039) | Bacteria | 2.20 | 2 | Oxidoreductases | protein-protein | m-m | ['Asymmetry: Hetero 2-mer - AB', 'BA1: Hetero 2-mer - AB', 'BA2: Hetero 4-mer - A2B2'] |
| 5V6P | 1 | AB | alpha | ER-associated protein degradation (ERAD) protein. Hrd1 channel in complex with Hrd3 | Saccharomyces cerevisiae | Eukaryota | 4.10 | 1 | Sec and Translocase Proteins | protein-protein | m-m | ['Asymmetry: A2', 'BA1: Homo 2-mer - A2'] |
| 5VRE | 1 | AC | alpha | TMEM175 lysosomal K⁺ channel | Chamaesiphon minutus | Eukaryota | 3.30 | 1 | Channels: Potassium, Sodium, & Proton Ion-Selective | protein-protein | multimer | ['Asymmetry: A4', 'BA1: Homo 4-mer - A4'] |
| 6BAA | 1 | AB | alpha | Cryo-EM structure of the pancreatic beta-cell KATP channel bound to ATP and glibenclamide | Rattus norvegicus | Eukaryota | 3.63 | 8 | Metal transporter | protein-protein | multimer | ['Asymmetry: Hetero 8-mer - A4B4', 'BA1: Hetero 8-mer - A4B4'] |
| 6F0U | 1 | AB | alpha | GLIC mutant E35A | Gloeobacter violaceus (strain PCC 7421) | Bacteria | 2.35 | 5 | Membrane protein | protein-protein | multimer | ['Asymmetry: Homo 5-mer - A5', 'BA1: Homo 5-mer - A5'] |
| 7AHL | 1 | AB | beta | α- hemolysin | Staphylococcus aureus | Bacteria | 1.90 | 1 | Adventitious Membrane Proteins: Beta-sheet Pore-forming | protein-protein | multimer | ['Asymmetry: A7', 'BA1: Homo 7-mer - A7'] |

| | Toxins/Attack Complexes |